# Commercializing metal production in the modern era

Charles Osarinmwian[1], Agatha Dutton[2], Josephine Osarinmwian[3]

[1]School of Chemical Engineering and Analytical Science, The University of Manchester, Oxford Road, Manchester M13 9PL, United Kingdom

[2] Salford Business School, University of Salford, Lady Hale Building, Salford M5 4WT, United Kingdom

[3]Nats Wholesale Ltd, 64 Derby Street, Greater Manchester, Manchester M8 8AT, United Kingdom

The dominant feature of industrial development in the nineteenth century was the use of power, but closely associated with it and of scarcely less importance was the enormous increase in the use of metals made possible by metallurgical progress (Hartley 1942). Here we review the development and use of three wonder metals, and propose next-generation electrochemical cell designs for metal production.

## Titanium

In nature, Titanium is found in igneous rocks where it forms components of acidic and basic magmas. The most useful mineral for the extraction of titanium and titanium compounds is rutile ($TiO_2$); although rarer than ilmenite ($FeTiO_3$) its titanium dioxide content is higher (Heinz et al. 2005). The industrial process for titanium production is the pyrometallurgical Kroll process (Fig 1), which was invented by William Kroll in Luxembourg in the 1930s. On the demand side, titanium is plentiful in the earth's crust and its alloys exhibit outstanding structural efficiency, biocompatibility and corrosion resistance for application in a wide range of industries. On the supply side, the selling price of titanium and its production costs need to be lowered. Limitations of the Kroll process have held back the widespread use of titanium where titanium prices have been 15 times higher than aluminum and 60 times higher than steel (Marsh 2010a). Also, the worldwide production and consumption of 40 million tons of aluminum and 1.2 billion tons of steel in 2009 dwarfed ~ 100 thousand tons of titanium (Marsh 2010b). This was compounded by the forecast that world titanium consumption will increase at an average annual rate of 4-10 % between 2011 and 2025 (Hogan et al. 2009).

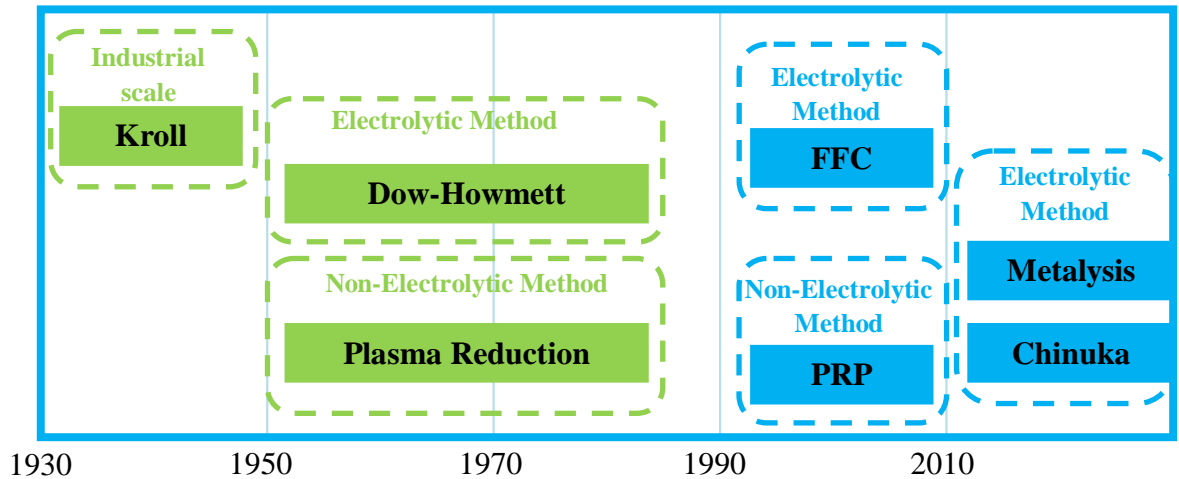

**Figure 1: History of titanium extraction. Although titanium is produced at an industrial scale by the Kroll process, significant advances in titanium production has evolved since its discovery by William Gregor in 1791.**

In the Materials Department at Cambridge University in 1997, Derek Fray, Tom Farthing, and George Chen made an unexpected and dramatic finding during their investigation into the elimination of an oxide film from the surface of titanium. They observed that it was possible to electrolyze solid oxide films on titanium foil by making the foil cathodic in a molten calcium chloride electrolyte (Chen et al. 2000; Flower et al. 2000). Under the influence of an applied cell voltage, oxygen ions travelled from a porous cathode towards a graphite anode through the molten salt. These ions discharged at the graphite anode and liberated as a mixture of oxygen and carbon oxide gases. This revolutionary discovery became known as the FFC Cambridge process; taking its name from the inventors and their University. This simple, environmentally friendly and cost effective process is best suited to converting high-melting transition metal oxides and actinides to metal (Abdelkader et al. 2013) as well as silica to silicon (Nohira et al. 2003).

## Aluminum

The molten salt electrochemical synthesis of Aluminum by the Hall-Héroult process has been well established for over a century (Rabinovich 2012). Alumina reduction is performed in a Hall-Héroult cell consisting of a steel shell (9-12 m long, 3-4 m wide and 1-1.5 m depth) lined with refractory alumina, carbon and a thermal insulator. The bottom of the cell is lined with carbon blocks inlaid with steel and current distributor bars that are preferably inlaid horizontally in order to generate a relatively uniform current distribution (Li et al. 2003). There are two types of consumable carbon-based anode: monolithic self-baking (Soderberg) and prebaked, where cells containing prebaked anodes are more efficient and differ only in fabrication and connection of the carbon anode (Jarrett 1987b; Prasad 2000). These anodes are consumed at a rate of ~ 2 cm day$^{-1}$ and maintain an anode-cathode gap of 4 to 5 cm.

Electric current enters the cell through the anode and flows through 3 to 6 cm of molten cryolite, which contains additives (e.g. $CaF_2$) and alumina, to the molten Aluminum deposited at the bottom of the cell. In commercial cells, the optimum current density is ~ 1 A cm$^{-2}$ (total current capacity range from 60 to 500 kA) that can produce > 450 to 4000 kg of Aluminum per day (Choate and Green 2003). Current efficiency is limited to 85-95 % at a cell voltage 4.0-4.5 V. Anode and cathode reactions create localized conditions that are different to bulk molten cryolite. The gas evolved from anode reactions generate bubbles, which lowers the effective molten cryolite conductivity and so increases electrical resistance:

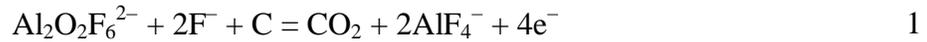

$$Al_2O_2F_6^{2-} + 2F^- + C = CO_2 + 2AlF_4^- + 4e^- \qquad\qquad 1$$

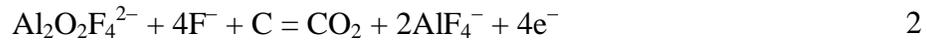

$$Al_2O_2F_4^{2-} + 4F^- + C = CO_2 + 2AlF_4^- + 4e^- \qquad\qquad 2$$

Also, the polarization effects at the cathode contribute much less to overpotential than at the anode. In addition, no gas bubbles, which influence both resistance and concentration polarization, are produced at the cathode:

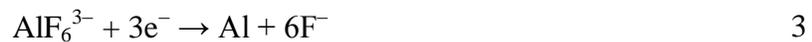

$$AlF_6^{3-} + 3e^- \rightarrow Al + 6F^- \qquad\qquad 3$$

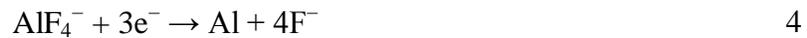

$$AlF_4^- + 3e^- \rightarrow Al + 4F^- \qquad\qquad 4$$

Controlling the operating variables of the cell becomes more critical as temperature is lowered even though changes in operating temperature have a minor effect on the theoretical energy requirements (Choate and Green 2003). Kasherman and Kazacos (1988) found that increasing the temperature increases the electrical conductivity of molten cryolite while reducing the bubble contribution to electrical resistivity at low anode-cathode gaps.

**Alumina Distribution.** The deposition of alumina at the cell bottom, caused by poor dissolution and diffusion of alumina in molten cryolite, creates alumina concentration gradients that generate uneven current density distributions. At very low alumina concentrations (< 2 wt.%) a phenomenon referred to as the anode effect, which is common in many molten salt processes, occurs (Haupin 1987a). Its physical manifestation is the growth of large bubbles on the horizontal underside of an anode, which indicates the decreased ability of the molten salt to wet the anode. These large bubbles have been known to coalesce, forming a single large bubble that covers most if not all of the anode underside (Haupin and McGrew 1975; Haupin 1987a). Rye et al. (1998) observed that when an anode effect is imminent, a redistribution of current among the anodes takes place; and that anodes located in regions where the concentration of dissolved alumina is low draw less current. In contrast, Cassayre et al. (2002) concluded that alumina concentration does not have a major influence on the size of gas bubbles.

Also, Zhou et al. (2007a) reported that in order to avoid alumina deposition at the cell bottom, alumina must be distributed evenly. They concluded that the dissolution and diffusion of alumina was mainly dependent on molten salt circulation, where anode gas bubbles were likely to be the main driving force of circulation. Zhou et al. (2007b) revealed that gas bubble velocities increased with increasing current density. Using multiple anodes, Chesonis and LaCamera (1990) investigated the influence of gas-driven bubble flow on alumina distribution. Much of the gas bubbles leaving the anode were periodically released into the space between the sides of adjacent anodes, where alumina distribution was dominated by gas-driven bubble flow. Moreau and Ziegler (1988) stipulated that relatively large channels between adjacent anodes govern molten salt circulation.

**Anode material.** Innovations in inert (or oxygen-evolving) electrode materials would improve cell efficiency. The relatively large gas bubbles at the horizontal underside of graphite anodes contributes to high ohmic potential drop and low electrical conductivity of molten salt in the bubble-free anode-cathode space (Zoric and Solheim 2000; Bech et al. 2001; Vogt and Kleinschrodt 2003). Attempts have been made to lower the 120-130° contact angle and large overpotentials by increasing the alumina concentration (Haupin 1987b) and doping the anode with lithium salt (Prasad 2000) respectively. However, the use graphite electrodes in such cells are unacceptable as their replacement is much more difficult and their consumption enlarges inter-electrode gaps that cannot be compensated by repositioning the electrodes as in conventional Hall-Héroult cells (Zoric and Solheim 2000).

Dimensionally stable inert Cu-Ni, Cu and $SnO_2$ anodes generate more uniform current distributions along horizontal undersides due to smooth release of more than 95 % smaller gas bubbles (where a bubble forms ~ 0° contact angle at the underside) from ~ 60 % thinner bubble layers than graphite anodes (Cassayre et al. 2002). Inert materials could also find use in wing-pattern stub designs, which minimize potential drop in the anode-stub connection area (Hou et al. 1995; Richard et al. 2001), and cathodic protection systems (Fray 2012; Osarinmwian 2013). The development of such inert anodes permit cell operation with a reduced anode-cathode gap, allow more efficient cell designs to be developed and significantly reduce process control problems such as cell disturbance from daily carbon anode changes. It is important to note that the development of a viable inert anode material is an extremely difficult task because of stringent requirements imposed (Sadoway 2001).

**Anode design.** In practice, flat horizontal anode undersides generate low rates of mass transport and tend to induce the anode effect (Vogt 2008). This leads to relatively high cell operating costs and highly non-uniform current distributions on electrode surfaces.

*Sloped Underside.* Thonstad (1984) studied the sensitivity of inclination angle of an anode underside with respect to a radial horizontal plane in an *aqueous* model of the Hall-Héroult process. It was found that anodes with flat undersides released gases in a pulsating manner while a 0.5° inclination angle established a smooth gas release. Troup et al. (2003) proposed an invention relating to bubble-driven flow to enhance the dissolution and distribution of alumina in a cell. They proposed that a sloped anode underside would induce bubble-driven circulation patterns in molten cryolite salt. A 0.5-3° inclination angle generated smooth bubble-driven flow and local circulation patterns in the direction of a slope. Also, grouping sloped anode undersides allowed control and enhancement of circulation patterns:

- 4 sloped horizontal anodes were orientated to generate a counter-clockwise circulation pattern in the molten salt;

- 6 sloped horizontal anodes were orientated to generate a clockwise circulation pattern at the left side and right side of a cell, with a counter-clockwise circulation pattern in the middle of a cell;

- 18 sloped horizontal anodes were orientated to generate a serpentine circulation pattern in the molten salt.

Zhou et al. (2007b) simulated the effect of changing the inclination angle from 1.5° to 10° on the velocity of gas bubbles in the anode-cathode gap. They found that the velocity of gas bubbles increased within this range of inclination angle.

*Undulated Underside.* Shekhar and Evans (1990) tested the performance of anode designs with grooved undersides using an *aqueous* model of the Hall-Héroult process. Tests were performed in a large electrolytic tank consisting of anodes suspended in horizontal or vertical configurations. Fine bubble dispersions, similar to those generated by inert anodes in the Hall-Héroult process, were induced as follows: forcing compressed air through porous graphite to simulate gas flow and adding butanol to water in the electrolytic tank. The velocity of gas bubbles was measured using a laser Doppler velocimeter for flat and grooved anode undersides. They found that the shape of the underside had a significant effect on circulation patterns in the anode-cathode region and that rapid circulation was obtained when the cell was operated with a fully submerged anode. In a similar experimental set-up, Shekhar and Evans (1994) later found that bubble velocity and electrical resistance measurements in the anode-cathode gap indicated superiority of grooved anode undersides over flat anode undersides. They predicted that grooved anode undersides would lower electrical resistance in the anode-cathode region by 40 % and minimize Aluminum re-oxidation by uniform molten cryolite circulation in the anode-cathode region. Also, they recommended that cells in the Hall-Héroult process should operate with grooved horizontal anodes.

However, Shekhar and Evans (1996) later observed that a high proportion of gas bubbles were not effectively removed from grooved anode undersides. They tackled this problem by improving groove design in the anode, and performing simulations based on an *aqueous* model of the Hall-Héroult process. They concluded that an energy saving of 2-2.5 kWh kg$^{-1}$ Aluminum may be achieved by installing there improved grooved horizontal anode into a cell. Lorentsen et al. (2005) proposed an invention of grooved anode designs for Aluminum production in a cell. There designs were intended to control the removal of gas bubbles as well as inducing well defined circulation patterns in the molten salt. They performed optimization studies on their anode designs and found:

- Anode-cathode gap can be minimized due to efficient gas bubble removal in the anode-cathode gap;

- Grooved slopes on an anode or group of anodes can be manipulated to generate particular circulation patterns in the molten salt;

- Total immersion of anodes in molten salt generate strong and controlled circulation in the molten salt;

- Smooth undulated undersides prevent localised regions of high current density.

*Perforated Underside.* Technological advances in cell design in the Hall-Héroult process are focused on minimizing operational costs while improving process efficiency with anode designs that efficiently drain gas bubbles away from an underside. De Nora (2002) designed a perforated anode that comprised a series of horizontally parallel rods that are spaced apart in a generally co-planar arrangement to form longitudinal flow-through openings. Simulations of voltage field distribution and molten salt circulation were performed around these parallel rods in a cell. It was shown that the geometry of the anode offered a large surface area and generated relatively uniform current distributions. The positioning of anode rods offered excellent gas release and an opportunity to evaluate the optimum depth of anode rods based on circulation rates in molten salt.

*Anode Stub.* The attainment of a uniform current distribution within an anode is mainly influenced by the contact area between the porous, electrochemically active anode face and the current feeder. Electrical contact and physical support is obtained through Aluminum or copper rods welded or bolted to steel stubs. The stubs are set in the anode sockets (i.e. anode-stub) where molten cast iron is poured around them to produce a strong joint with low electrical resistance (Jarret 1987b). Hou et al. (1995) and Richard et al. (2001) concluded that the high contact surface area offered by the wing-pattern stub would minimize the potential drop in the anode.

## Tantalum

In mid-2008, tantalum consumption was on a strong growth trend in its principal markets of electronics and aerospace super-alloys, which made up ~ 75 % of tantalum consumption (Buetow 2009). However, the global economic downturn had a marked effect on the tantalum supply/demand balance. Several tantalum-based projects were at various stages of development around the world in 2009. Success in these projects could resolve the tantalum supply issue for the foreseeable future and obviate the need to source conflict minerals. Carbone (2009) reported that the global market for tantalum capacitors will decline from ~ £1.5 billion in 2008 to ~ £1.4 billion in 2010 and forecasted that it will increase back to ~ £1.5 billion by 2013.

The demand for tantalum capacitors was weak due to the overall industry downturn with more electronics equipment using ceramic capacitors over tantalum. However, Wireless News (2010) reported that the unique properties of tantalum are indispensable to a few applications. Increased usage of electronic equipment such as computers, mobile telephones and video cameras, is expected to boost demand for tantalum capacitors. Furthermore, research activities for developing new applications for tantalum are underway. For example, AVX developed a range of high-capacitance PulseCap tantalum capacitors designed for applications that require bulk capacitance to boost transmitter power, such as wireless cards and smart meters (Electronics Weekly 2010).

### Electrochemical cell designs

**Governing equations.** The prediction of current distribution is an essential step in the rational design and scale-up of electrochemical cells (Armstrong et al. 1968; Walsh 1991). The current density at a given location in a cell domain $\mathbf{J}$ is related to the electric field strength $\mathbf{E}$ at the same location as the current density in a stationary coordinate system using Ohm's law $\mathbf{J} = \kappa \mathbf{E}$ where $\kappa$ is the effective electrical conductivity of molten salt. The electric field strength is defined as the force per unit charge on a stationary test charge $q$ by $\mathbf{E} = \mathbf{F}/q$ where $\mathbf{F}$ is the electric force experienced by $q$. The $\mathbf{E}$ is related to the charges generated by Gauss' law. The electromagnetic interaction between a moving positive $q$ and an electromagnetic field is the source of $\mathbf{F}$ according to the Lorentz force law $\mathbf{F} = q[\mathbf{E} + (\mathbf{v} \times \mathbf{B})]$ where $\mathbf{v} \times \mathbf{B}$ is the vector cross product of the instantaneous velocity of a positive test charge and the magnetic field strength.

**Finite element method.** The finite element method is used to solve governing equations. Governing equations are expressed as integral equations and partial differential equations that are readily represented and approximated by the finite element method (Zimmerman 2006). Most partial differential equations encountered in science and engineering are second order because the principle of minimum Fisher information introduces a second order operator of a field quantity as the highest order term. The method uses a weak formulation of the constraints of field variables in coupled systems of partial differential equations to make discontinuities integrable. This is very important at points of discontinuity in a cell domain. Equivalence between a partial differential equation and its weak form is considered at steady state according to non-linear functions of a single dependent field variable that is multiplied by an arbitrary test function and integrated (Zimmerman 2006).

**Convergence, accuracy and dynamics.** The number of finite elements within a cell domain is directly related to the accuracy of numerical solution **u**. Convergence to **u** is monitored by an implicit error criterion which allows error distribution within a finite element mesh to be controlled by changing the size of finite elements. Given that refinement techniques based on local mesh refinement are most commonly used (Zhu et al. 1993), each finite element in the region of highest error in the current density field is either subdivided into four equally shaped finite elements according to the regular split method in 2D models or divided by the longest edge into two new elements according to the longest bisection method in 3D models. Given that there are > 100 types of finite elements in use, triangular elements are generally used for 2D problems (Zhu and Zienkiewicz 1988; Zhu et al. 1993) with tetrahedral elements in 3D problems (Lo 1998; Merrouche et al. 1998); the accurate solution criterion for 2D and 3D problems are $\leq 5$ % and $\leq 10$ % respectively. Although a solution completely independent of finite element size is desirable, mesh independent solutions are unlikely to be achievable due to large computational effort and geometric singularities.

The convergence rate is significantly influenced by the mesh refinement technique in a cell domain containing geometric singularities. The *p*-version of the adaptive refinement technique (i.e. refinement by adding successive high order finite element interpolation basis functions) and a combination of both *h*- and *p*-versions may be required because they are well suited to handling singularities (Zhu et al. 1993). Also, special finite elements containing singular interpolation functions with the right type of derivatives (e.g. modified interpolation functions with shifted nodal points) could avoid considerable instability in convergence at geometric singularities. Lack of symmetry or dimensional changes in a cell domain requires multiple mesh sensitivity tests. However, such tests would not be worth the additional computational effort of using a fine mesh for a small improvement in results where gradients are located in the same regions and of the same order of magnitude in different cell configurations. The concept of mesh refinement can be applied to reducing the time step so that the implicit time integration scheme represents accurately the dynamic behavior of transient phenomena in a cell.

The Jacobian matrix $L(\mathbf{u_0})$ describes the orientation of a tangent plane to a function at a given point within a cell domain and represents the best linear approximation to a differentiable function near that point; whereas the stiffness matrix $K(\mathbf{u_0})$ characterizes the local stability of the time-dependent evolution of nonlinear dynamical processes governed by partial differential equations (Zimmerman 2006); the stiffness matrix is the negative Jacobian. Improved approximations of $\mathbf{u}$ are found by iteration until $\mathbf{u}$ is an acceptably accurate solution. For time-dependent evolution, COMSOL Multiphysics uses the DASPK version of a differential-algebraic equation solver to integrate over time step increments such that the implicit time integration scheme accurately describes time-dependent behavior. An initial approximate $\mathbf{u_0}(t)$ for the new approximate $\mathbf{u}(t)$ is developed by evaluating a predictor polynomial, which interpolates approximates at previous time steps. The new approximate is then computed in a corrector step by solving a nonlinear system of equations using Newton iteration at each time step.

**Results.** Figure 2 shows an electrochemical cell with 2D axi-symmetry. Imposing boundary conditions: −3.2 V (current collector/metal oxide), grounded (anode/melt) and insulated (current collector/melt), caused electric field lines to move according to the Lorentz force law while the strength of equipotential field lines, which pass radially and perpendicular to electric field lines, diminish away from discontinuous points on the current collector/metal oxide interface according to Coulomb's inverse square law.

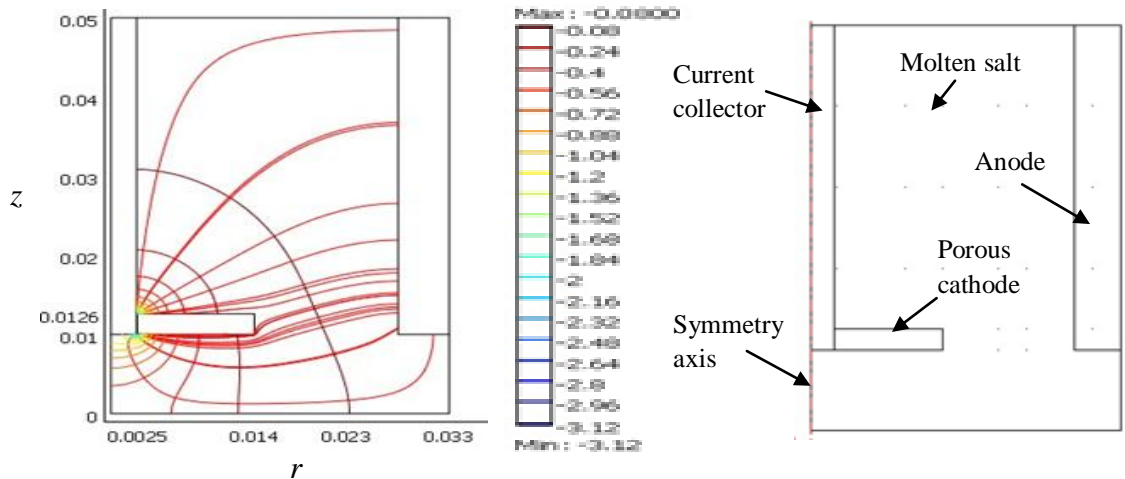

**Figure 2: Concentric cylinder electrochemical cell design containing a *Standard* porous metal oxide cathode.**

*Close* and *Single rod* electrochemical cells generate the most uniform current density distribution (i.e. lowest standard deviation in current density) along the current collector/metal oxide interface (Fig 3). *Close* consists a central cathode surrounded by anodes where *Loose* consists a central anode surrounded by cathodes. In the rod-based designs, parallel electrode rods would concentrate ~ 16 % more current density at the distance of closest approach to an electrode surface than a parallel plate (Prentice 1991).

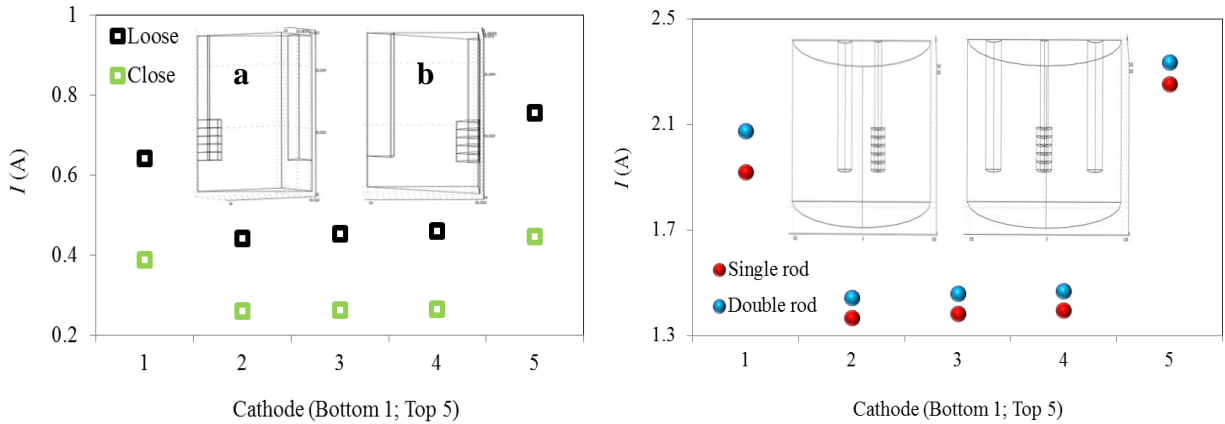

**Figure 3: Current distribution in electrochemical cell designs containing a stack of 5 porous cathodes and vertical anode rods (a, Close; b, Loose).**

Current distribution modeling in Hall-Héroult cells has played an important role in the calculation of heat, magnetic, force and flow fields (Johnson 1988; Feng et al. 1990). The shape of anode undersides affects the potential and current distribution (Fig 4; Osarinmwian 2014a).

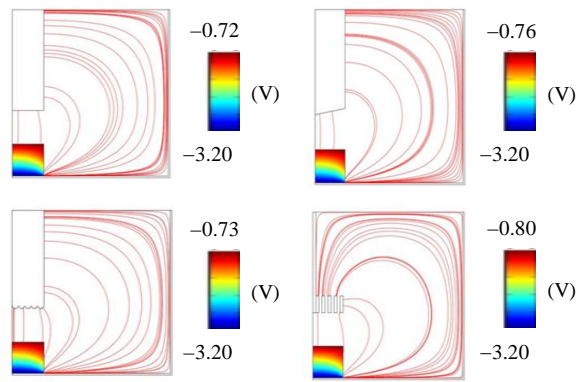

**Figure 4: Potential distribution in electrochemical cell designs containing shaped horizontal anode undersides.**

Near-slit designs are favorable because they generate the most uniform current density distribution (Fig 5). Production of porous near-net-shaped metal products would facilitate diversification, adaption and even reinvention of product designs to match evolving market conditions in high-tech sectors. In these novel designs, future work would involve observing three-phase interline movement (Osarinmwian 2014b), and studying the effect of pore shape and orientation on thermal properties (Bari et al. 2013). It is important to note that census data shown that U.S. and foreign produced titanium prices varied between 2003 and 2012 because of the relationship between product design, operating costs and titanium production capabilities in different countries (see Osarinmwian 2012).

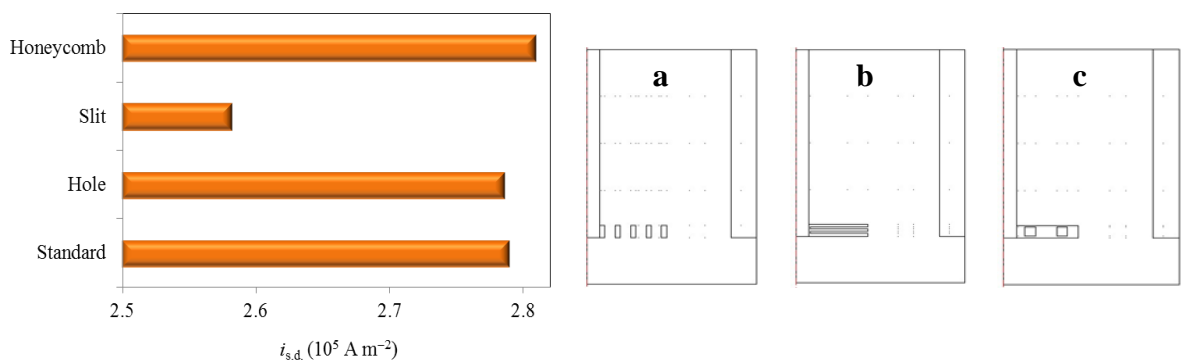

**Figure 5: Uniformity of current density distribution along the current collector/metal oxide interface in electrochemical cell designs containing near-net-shaped porous metal oxide cathodes (a, Honeycomb; b, Slit; c, Hole).**

Many Group IV elements in the Periodic table are able to dissolve substantial quantities of oxygen; most other metals have negligible solubility for oxygen (Fray 2006). For example, using the FFC Cambridge process to convert chromium oxide to chromium in molten calcium chloride takes 2-6 h because time is not heavily dependent on oxygen ion diffusion out of porous phases. Also, uranium oxide to uranium occurs at a temperature close to its melting point. Therefore, uranium tends to be dense with closed pores that blocks ingress of molten calcium chloride.

## Future directions: Monopolar vs bipolar

In a concentric cylinder electrode design, an anode annulus with an inwardly-facing porous surface encloses a solid cathode bar at its centre; the surface area of the porous anode surface is at least 5 times greater than the solid cathode surface (De Nora 1994). This design allows gas bubbles to be released through the upper end of the anode while the melt is circulated through the lower end (openings in the anode wall located below the level of the melt allow circulation) in which the effect of bubble layer resistance on current density is minimized by minimizing the anode immersion depth. The uniformity in current distribution in a cell can be optimized by reducing the thickness of the anode wall and tapering the cathode cross-sectional area from top to bottom (Secrist et al. 1986). These novel inventions offer significant benefits over conventional Hall-Héroult cells:

- The ratio of the active surface area of tubular anode to cathode rod can be set at any convenient value;

- The use of vertical electrodes promotes rapid escape of gas bubbles, which allows minimisation of the anode-cathode gap;

- The design of electrodes are very simple and their vertical installation permits easy electrode replacement during operation;

- The thickness of the tubular anode wall can be chosen in order to change the current density in the anode for optimization studies and to lower material costs;

- The construction, maintenance and operation costs of the cell are lowered considerably.

De Nora and Sekhar (2003) proposed an invention for the application of multi-polar connections in a cell containing vertically inclined electrodes. A series of anode plates inclined at 45° above shaped cathodes formed a grid-like assembly. Apertures were incorporated in the upper parts of anode plates to facilitate the release of gas bubbles. The cells described meet the need for chemical and thermal stability, ease of fabrication and construction, and fundamental process information required for developing an industrial scale process. However, attempting to use these cells for large scale operation is not feasible due to labour intensive requirements, limited production per unit volume, and lack of flowing molten salt.

In recent years there has been an increasing interest in the use of bipolar cells for fused salt electrolysis for light metal production. The bipolar electrolyser requires the external current leads to be connected only to the two terminal electrodes, simplifying the electrical circuitry and enabling the electrolyser to be more compact design with minimal footprint. This energy efficient cell type operates without busbars within an individual cell stack while generating more uniform current distributions over electrode surfaces at lower current (and higher voltage) than its monopolar analogue (Kodym et al. 2007). In molten salt electrochemistry, the use of an alkali metal bipolar electrode in the electrodeposition of Ti from molten chloride salts (Maja et al. 1990) and bipolar electrode cells for electrorefining of scrap Al (Ueda et al. 2001) has been investigated. Most notable, bipolar technology has found use in Al production (Alcoa cell and Hall-Héroult cell) and Mg production (Alcan-Ishizuka cell). For instance, Alcoa cells produced an equivalent amount Al to monopolar cells where increasing the number of bipolar compartments (i.e. forming more inter-electrode channels) increased Al production (Jarrett 1987a).

Although bipolar cells offer large scale operation, their design must meet the requirements associated with the supply, circulation and removal of products (including heat). Ionic conduction paths in molten salt between neighbouring bipolar electrodes and a conducting cell wall lead to the existence of bypass current and a subsequent loss of current efficiency. This problem could be minimized by increasing the cell current and height of the bipolar cell, bounding bipolar electrodes with an upwardly extending rim around the electrode edge, and electrically insulating the internal wall of the cell (Feng et al. 1990; Dudley and Wright 2014). It is important to note that it is possible to obtain detailed information about the influence of the bypass current flow on the current density distribution at the edges of bipolar electrodes by using the Ishikawa-Konda approach (Rousar and Thonstad 1994).

A ground-breaking bipolar cell design for metal powder production recommended coupling a molten salt reservoir (comprising a filtration system) with a bipolar cell housing such that molten salt continually circulates through the cell (Rao et al. 2013; Dudley and Wright 2014). Considering molten calcium chloride, continual re-circulation would increasingly raise the current efficiency (i.e. lower electronic background current) since the passage of purified molten calcium chloride would increasingly lower calcium oxide concentrations (c.f. double-melt electrolysis, Chen et al. 1999). Another innovation could involve rotating a bipolar electrode through different potential regions (Nadebaum and Fahidy 1973). It is important to note that within a bipolar cell stack a number of requirements must be simultaneously satisfied where some requirements are in opposition. Also, the business operations (Osarinmwian 2014), ease of maintenance, product removal, health and safety, and reliability are key factors required to reach an economic minimum cost.

**Dedication**

This work is dedicated to Mama Agatha Adesode Erhumwunse who was born on the 5th of February 1912. Married to the late Pa Paul Omokaro Erhunmrnwnse she was blessed with 7 children: Miss Anna Erhunmrnwnse, Mrs Josephine Osarinmwian, Mrs Mather Udubor, The late Mr Jerome Erhunmrnwnse, Mrs Margaret Dutton, The late Mrs Theresa Akimus, Mr Emanu Erhunmrnwnse.

Mama A. A. Erhumwunse was an excellent spiritual singer and she inspired people to love God. Her favorite songs included: Jerusalem my Happy Home; Mary must be Honored in my Life; Ave Maria; Abide with Me; All things Bright and Beautiful.

Mama A. A. Erhumwunse was born and bred in a catholic home and professed the catholic faith all her life. As a teacher she taught and was head mistress in various Schools in Benin City (Edo State, Nigeria) (e.g. Benedict Catholic School formerly Iyoba Primary School, and Ediaken Primary School) for 42 years before retiring in November 1983.

Mama A. A. Erhumwunse was an active member of the National Union of Teachers, Christian Women's Organization, St Anthony's Catholic Women Society, and was a co-founder of the St Patrick Church (Ogbowo, Benin City). She was bestowed the prestigious *Papal Medal* by his holiness The late Pope John Paul II to the Glory of God.

Now Mama A. A. Erhumwunse' legacy lives on in her children, grandchildren, great grandchildren and so on. She will be loved eternally by all. May her gentle soul rest in peace forever AMEN.